 \documentclass[aps,pre,twocolumn,groupedaddress, amsmath, amssymb, showpacs]{revtex4-1}
 
\newcommand{\MB}{\textrm{MB}}

\newcommand{\EP}{\textrm{EP}}
\newcommand{\MD}{\textrm{MD}}

\newcommand{\vc}{\mathbf}
\def\be{\begin{eqnarray}}
\def\ee{\end{eqnarray}}
\def\ben{\begin{eqnarray*}}
\def\een{\end{eqnarray*}}

\usepackage{graphicx}
\usepackage{color}

\begin{document}


\title{Modified Enskog Kinetic Theory for Strongly Coupled Plasmas}


\author{Scott D.\ Baalrud,$^{1}$ J\'{e}r\^{o}me Daligault$^{2}$}

\affiliation{$^1$Department of Physics and Astronomy, University of Iowa, Iowa City, IA 52242}
\affiliation{$^2$Theoretical Division, Los Alamos National Laboratory, Los Alamos, New Mexico 87545}


\date{\today}

\begin{abstract}

Concepts underlying the Enskog kinetic theory of hard-spheres are applied to include short-range correlation effects in a model for transport coefficients of strongly coupled plasmas.
The approach is based on an extension of the effective potential transport theory [S.~D.~Baalrud and J.~Daligault, Phys.~Rev.~Lett.~{\bf 110}, 235001 (2013)] to include an exclusion radius surrounding individual charged particles that is associated with Coulomb repulsion.
This is obtained by analogy with the finite size of hard spheres in Enskog's theory.
Predictions for the self-diffusion and shear viscosity coefficients of the one-component plasma are tested against molecular dynamics simulations. The theory is found to accurately capture the kinetic contributions to the transport coefficients, but not the potential contributions that arise at very strong coupling ($\Gamma \gtrsim 30$). Considerations related to a first-principles generalization of Enskog's kinetic equation to continuous potentials are also discussed.  

\end{abstract}




\maketitle


\section{Introduction\label{sec:intro}}

Plasmas found in several modern research areas can span a broad range of Coulomb coupling strengths. Examples include dusty, non-neutral, and ultracold plasma experiments, as well as dense plasmas found in inertial confinement fusion, high-intensity laser-matter interaction experiments and dense astrophysical objects. A research need common to all of these areas is a transport theory that is versatile enough to cover a broad range of coupling strengths, and can be evaluated efficiently enough to be incorporated into the fluid codes used to simulate these systems. To date, no systematic theory is available to do this, yet a workable theory may be obtained through ad hoc extensions of theories of simplified systems. 

We recently proposed an approach based on an effective interaction potential in an effort to extend conventional plasma transport theory into the strongly coupled regime~\cite{baal:13}. This is a physically-motivated approach based on a Boltzmann-like binary collision operator, but where many-body correlation effects are modeled through an effective interaction potential. By comparing with molecular dynamics (MD) simulations of diffusion and shear viscosity of the one-component plasma (OCP), this was shown to be successful at extending the binary collision approach well into the strongly coupled regime. However, one persistent feature was a 30-40\% underestimation of the collision rate in the range $1 \lesssim \Gamma \lesssim 30$, where $\Gamma \equiv e^2/(ak_BT)$ is the Coulomb coupling parameter, $a=(3/4\pi n)^{1/3}$ is the Wigner-Seitz radius and $T$ the temperature~\cite{baal:14,bezn:14}. The theory begins to break down at larger coupling strengths $(\Gamma \gtrsim 30$) where direct interaction contributions that are not included in a Boltzmann-like treatment begin to dominate. In this paper, we attempt to improve the theory by invoking ideas underlying the kinetic theory proposed by Enskog to extend the kinetic theory of Boltzmann to dense gases~\cite{ensk:22,ferz:72}.
The Enskog kinetic equation was developed for hard-spheres only and involves the introduction of corrections to the Boltzmann equation that account for the finite particle size.
Although no one has yet succeeded in deriving a similar equation for continous potentials, the concepts underlying Enskog's theory provide a valuable aid in understanding the physical origin of the underestimation of the collision rate, as well as a source of ideas to improve the effective potential theory.

Two basic assumptions limit the Boltzmann equation to dilute systems: (i) only pairs of particles collide simultaneously (i.e., binary collisions), and (ii) the ``Stosszahlansatz'' or ``molecular-chaos'' assumption. The effective potential theory relaxes assumption (i) somewhat by treating binary scatterers as interacting through the potential of mean force, rather than the bare Coulomb potential. The potential of mean force is obtained by taking the two scattering particles at fixed positions and averaging over the positions of all other particles~\cite{hill:60}. It is related to the pair distribution function by $\phi(r)/k_BT = -\ln [g(r)]$, and includes many-body effects of the background including screening and correlations. The present work shows that further extension may be realized by addressing assumption (ii) through Enskog's equation. 

\begin{figure}[b]
\includegraphics[width=7.0cm]{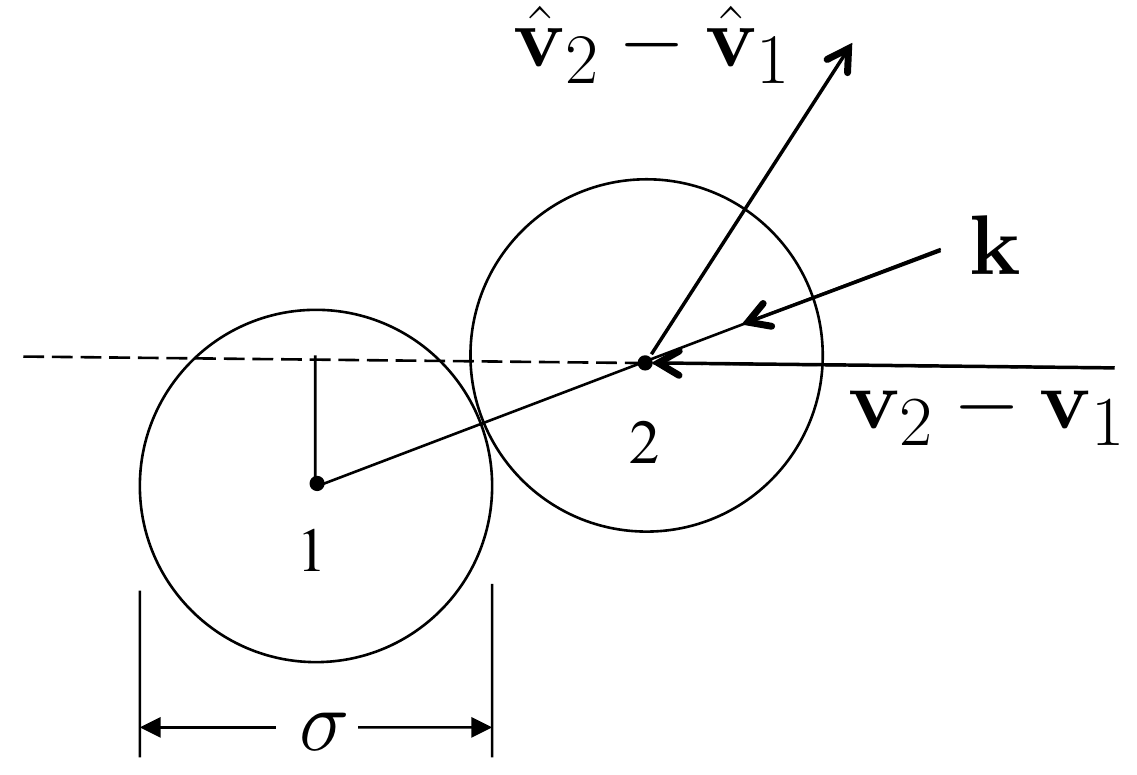}
\caption{Geometry of two colliding hard spheres of diameter $\sigma$ at the point of contact. }
\label{fg:spheres}
\end{figure}

Boltzmann's molecular chaos approximation assumes that particles are uncorrelated prior to a collision, and even ignores the difference in the positions of the two colliding particles by setting $\vc{r}_1 = \vc{r}_2 = \vc{r}$. Enskog's theory relaxes these assumptions by modeling molecules as hard spheres of finite diameter $\sigma$; see Fig.~\ref{fg:spheres}.  This treats the fact that particle centers cannot be closer than their physical diameter (an exclusion radius), which introduces both an aspect of correlation in the distribution of initial positions of scattering particles, as well as nonlocal aspects related to the scale at which momentum and energy transfer occurs in a collision. These give rise to qualitatively new features in the resulting fluid theory, including a non-ideal equation of state, and potential energy contributions to transport coefficients such as shear viscosity and thermal conductivity~\cite{ferz:72}. 
 
\begin{figure}
\includegraphics[width=7.8cm]{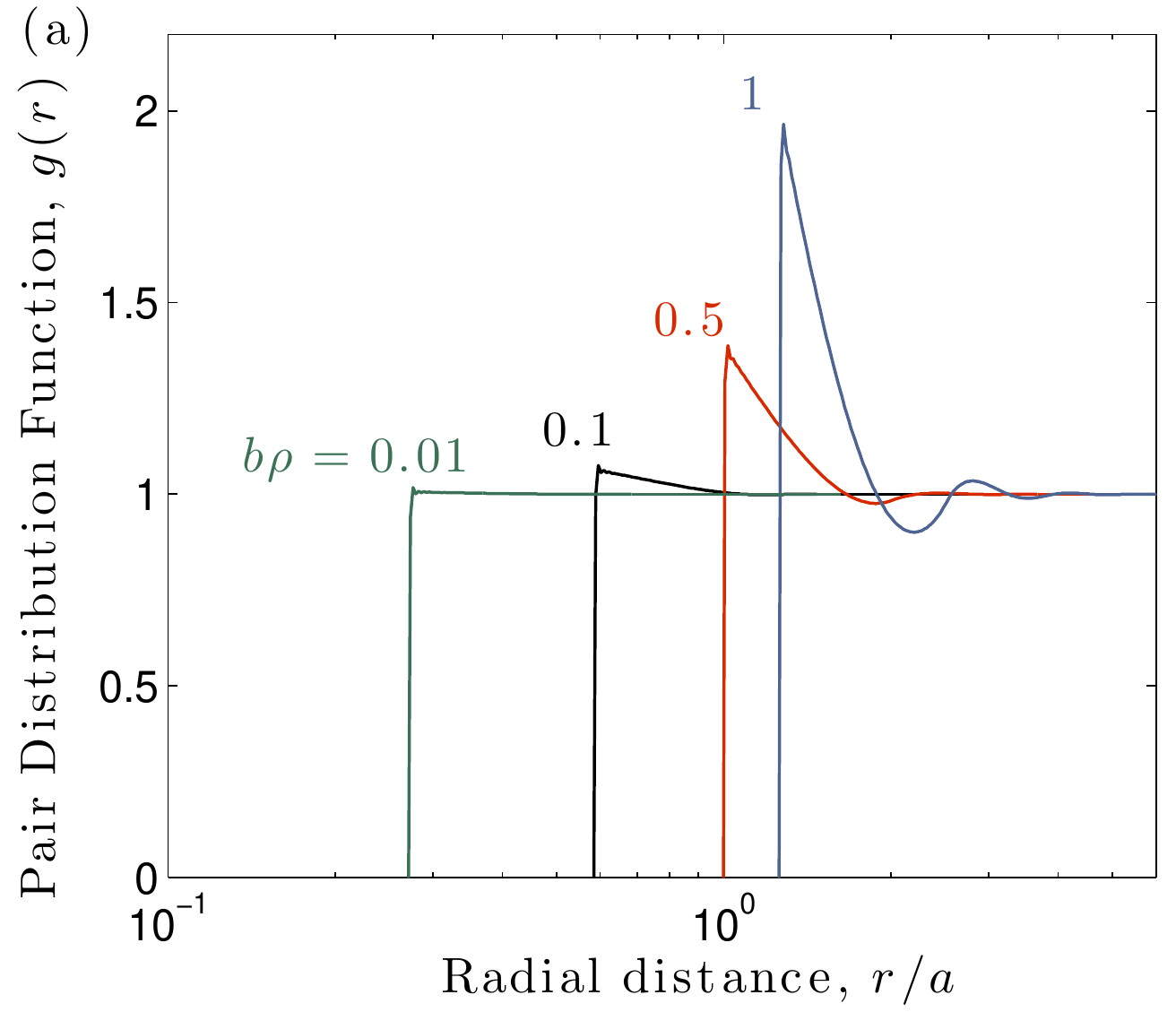}
\includegraphics[width=8.0cm]{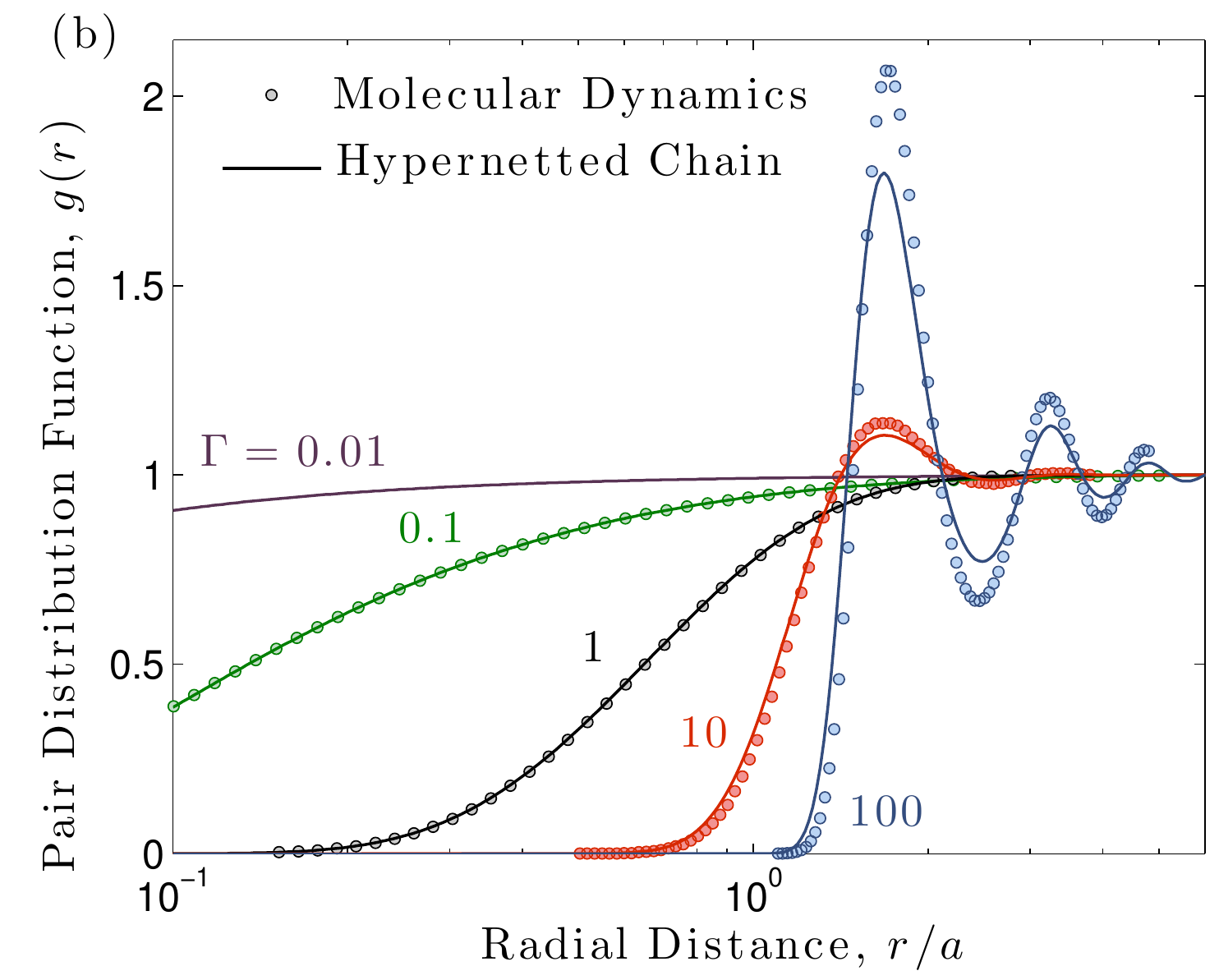}
\caption{(color online) (a) Pair distribution function for hard spheres calculated from the Percus-Yevick approximation.  (b) Pair distribution function the OCP calculated from the HNC approximation (lines) and MD simulations (circles).}
\label{fg:gr}
\end{figure}

Enskog's equation provides great insight, but real gases are not comprised of hard spheres. The hard sphere approximation is central in a systematic derivation of Enskog's equation~\cite{seng:61}, but successful descriptions of realistic fluids have been realized through \emph{ad hoc} identification of appropriate effective particle diameters~\cite{hanl:72}. The present work seeks a parallel approach for plasmas. At the outset, it is not immediately apparent that this is possible since the Coulomb potential is ``soft'' in the sense that it falls off gradually ($\propto 1/r$) from particle centers. A critical aspect of the present work is the effective potential concept: colliding pairs interact through the potential of mean force rather than the bare Coulomb potential. At strong coupling many-body correlations cause the potential of mean force to sharply fall-off at a fixed distance from the particle center, forming a ``Coulomb hole''; see Fig.~\ref{fg:gr}. We show that this can be associated with an effective particle radius, opening a path for applying Enskog's theory. 

Two parameters arise in Enskog's modification to the dilute gas transport coefficients: An increased collision frequency $(\chi)$ and the hard sphere packing fraction ($\bar{\eta}$). There is no systematic way to map these hard sphere parameters onto plasmas. We first determine them by equating MD simulations of the OCP self diffusion and viscosity coefficients with Enskog's expressions, taking the effective potential theory as the dilute gas component. Essentially no modification from the dilute gas effective potential theory is found for $\Gamma \lesssim 1$. The main modification in the region $1\lesssim \Gamma \lesssim 30$ is an increased collision frequency of approximately $30-40\%$ ($\chi \simeq 1.3-1.4$). Several attempts to model these parameters directly from the pair distribution function $g(r)$ are described. Although attempts at a systematic derivation were largely unable to quantitatively predict the Enskog parameters, a heuristic method based on associating the particle size with the Coulomb hole in $g(r)$ gives strong support for the notion of an increased collision frequency associated with an exclusion radius. 

The practical advantage of this approach is that it provides a computationally efficient and accurate way to determine transport coefficients in complicated systems. In comparison, MD simulations are known to provide highly accurate results, but are sufficiently computationally expensive that the prospect of building tables of transport coefficients over a large range of densities, temperatures, and mixture concentrations quickly becomes impractical. In this paper, the theory is validated against MD simulations of simple OCP and Yukawa OCP systems. Applications to more realistic systems such as mixtures and dense plasmas will be published subsequently. 

The remainder of this paper is organized as follows. Section~\ref{sec:enskog} briefly reviews the molecular chaos assumption in Boltzmann's theory, and Enskog's kinetic theory for the dense hard sphere gas.
The simplification provided by the hard-sphere approximation is that binary collisions take place instantaneously at known relative positions.
This is not the case with continuos interparticle positions and the difficulties that arise in a systematic generalization of Enskog's kinetic theory to treat particles interacting through continuous potentials are discussed in Sec.~\ref{sec:mensk}.
This section also introduces a modified Enskog transport theory based on an effective particle diameter.
Section~\ref{sec:size} provides a calculation of the Enskog parameters from MD simulations, and compares the results with attempts to determines these directly from $g(r)$.
Section~\ref{sec:md} provides a comparison between the previous effective potential theory calculations, calculations based on the modified Enskog theory, and MD simulation data for self-diffusion and shear viscosity of the OCP. Implications of these results are discussed in Sec.~\ref{sec:disc}. 

\section{Enskog Theory\label{sec:enskog}}

\subsection{Boltzmann's Stosszahlansatz} 

\begin{figure}
\centering
\includegraphics[width=8cm]{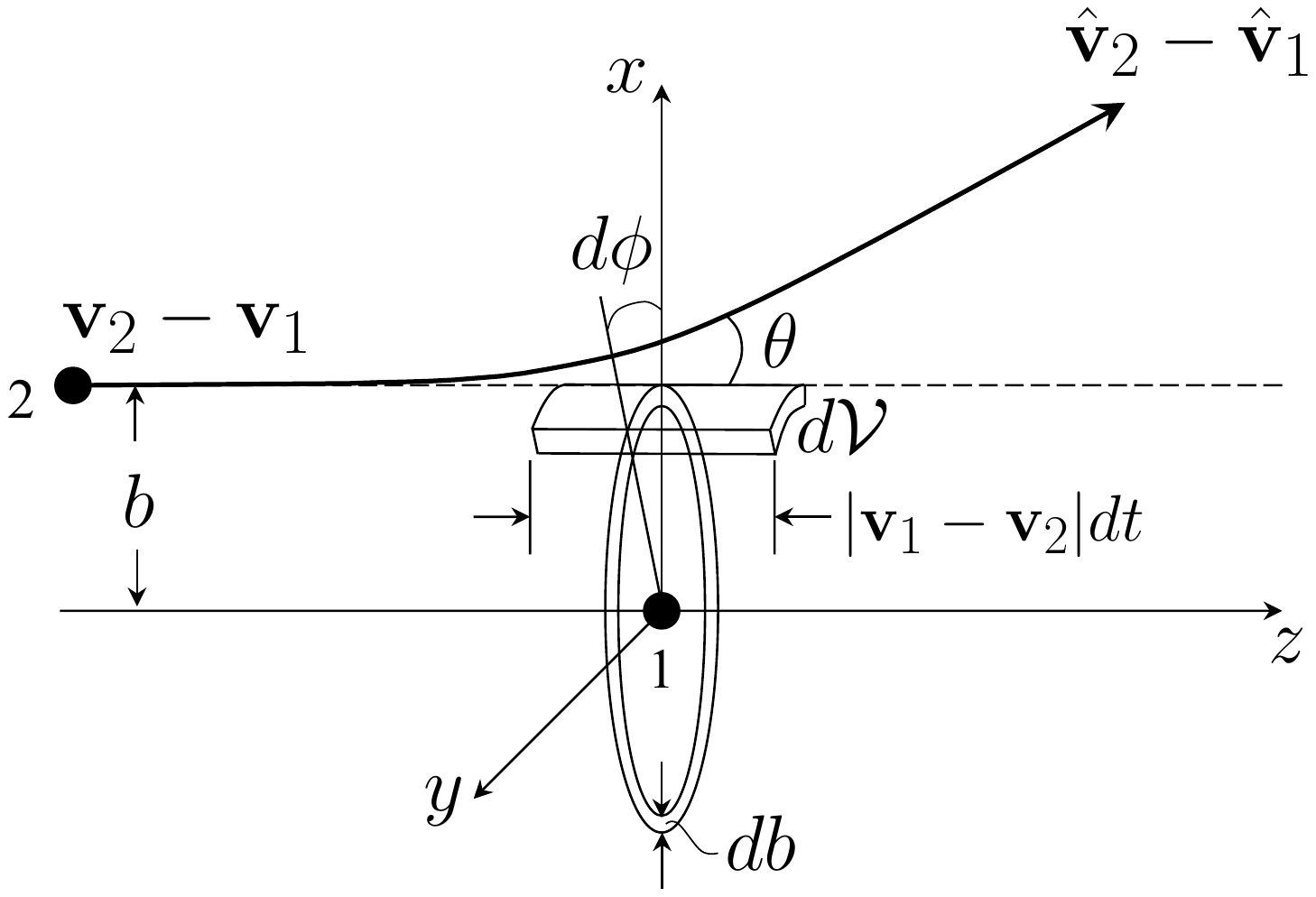}
\caption{\label{figure_binary_collision} Geometry of a binary collision between point particles in the center of mass frame.
}
\end{figure}

In the traditional derivation of the Boltzmann equation, one calculates the expected number of binary collisions $dN$ experienced during a small time interval $dt$ by particles `1' located in the phase-space volume element $d^3{\bf r}_1d^3{\bf v}_1$ centered about the phase-space point $({\bf r}_1,{\bf v}_1)$~\cite{ferz:72}.
For such a collision to occur, the colliding partners `2' with velocity in the range $d{\bf v}_2$ around ${\bf v}_2$ must initially lie within a cylinder, the so-called collision cylinder, having an area $b\,db\,d\phi$ and generator $-({\bf v}_1-{\bf v}_2)dt$. Its volume is $d{\cal{V}}=|{\bf v}_1-{\bf v}_2|b\,db\,d\phi\,dt$; see Fig.~\ref{figure_binary_collision}. Here, $b$ is the impact parameter and $\phi$ is the azimuthal angle.
In general, the expected number of binary collisions $dN$ is the integral over the collision cylinder of the total number of pairs of collision partners
\begin{equation}
dN=\int_{{\bf r}_2\in d{\cal{V}}}{f_2({\bf r}_1,{\bf v}_1;{\bf r}_2,{\bf v}_2;t)d{\bf r}_2d{\bf v}_2 d{\bf r}_1d{\bf v}_1}\,, \label{molecular_chaos}
\end{equation}
where $f_2$ is the phase-space pair distribution function.
The Boltzmann equation additionally relies on the molecular chaos assumption (a.k.a. Stosszahlansatz) according to which (i) colliding particles are entirely uncorrelated in position before colliding, i.e. $f_2({\bf r}_1,{\bf v}_1;{\bf r}_2,{\bf v}_2;t)\approx f_1({\bf r}_1,{\bf v}_1;t) f_1({\bf r}_2,{\bf v}_2; t)$, and (ii) the difference in position between two molecules is ignored and the distribution functions $f_1$ are evaluated at the same point $\vc{r}_1 = \vc{r}_2 = \vc{r}$ in space.
The molecular chaos approximation can then be written
\begin{equation}
f_2 (\vc{r}_1, \vc{v}_1; \vc{r}_2, \vc{v}_2; t) \approx f_1 (\vc{r}, \vc{v}_1, t) f_1 (\vc{r}, \vc{v}_2, t) \,,\label{eq:mc1}
\end{equation}
where $f_1$ is the single-particle distribution function.
Combining this with Eq.~(\ref{molecular_chaos}) yields 
\ben
dN \approx f_1({\bf r},{\bf v}_1,t) f_1({\bf r},{\bf v}_2,t) |{\bf v}_1-{\bf v}_2| b\,db\,d\phi\,dt d^3{\bf v}_2 d^3{\bf r}d^3{\bf v}_1 .
\een
The molecular chaos assumption restricts the validity of the Boltzmann equation to low densities, or in the context of the present paper, to small coupling strengths.

In general, particle positions are correlated and correlations increase with density or coupling strength.
For instance, in thermal equilibrium 
\be
f_2({\bf r}_1,{\bf v}_1;{\bf r}_2,{\bf v}_2;t)=g({\bf r}_1-{\bf r}_2) f_{\MB}({\bf v}_1) f_{\MB}({\bf v}_2)
\ee
where $f_{\MB}$ is the Maxwell-Boltzmann distribution and $g$ is the pair distribution function.
Physically, $ng(r)$ represents the radial density around a given particle.
In essence, molecular chaos assumes that $g(r)=1$ for all distances $r$.
Figure~\ref{fg:gr}b shows the pair distribution function of the OCP at several coupling strengths over a range of distances on the order of the interparticle spacing.
For weak coupling ($\Gamma \lesssim 0.1$), $g(r)$ is indeed very close to one over the range relevant to colliding particles. 
However, as the coupling strength increases, $g(r)$ transitions from 0 to $1$ at a radial distance on the order of the interparticle spacing. It stays nearly equal to zero over an increasingly large range as $\Gamma$ increases and then oscillates around one. 
The ``hole'' in $g(r)$ at small $r$ can be interpreted as an exclusion volume around each particle, which originates from the strong repulsion at small distances that the vast majority of particles can not penetrate due to their low kinetic energy.
Figure~\ref{fg:gr}a illustrates that a similar exclusion zone is found for hard spheres, which naturally arises from the finite size of particles.
In this case, the exclusion zone is simply related to the size of a particle and is independent of temperature and density; the value of $g(r)$ depends on the physical conditions only beyond the particle diameter ($r\geq\sigma$).
On the contrary, with particles interacting via a continuous potential, the exclusion zone depends on the physical conditions and cannot be delineated as clearly.

\subsection{Enskog's Equation} 

In 1922 Enskog was able to successfully abandon the molecular chaos approximation for hard sphere gases~\cite{ensk:22}.
He replaced it by an ansatz that includes effects of spatial correlations by invoking the following arguments: (i) Because of their finite size, the centers of colliding hard spheres are not at the same position at contact. Rather, they are separated by a distance equal to the particle diameter. Thus, $f_1({\bf r},{\bf v}_1,t) f_1({\bf r},{\bf v}_2,t)$ should be replaced by
\be
f_1({\bf r},{\bf v}_1,t) f_1({\bf r}-\sigma{\bf k},{\bf v}_2,t) \label{molecular_chaos_2}
\ee
in the molecular chaos approximation, Eq.~(\ref{eq:mc1}). Here, ${\bf k}$ is the unit vector joining the hard-sphere centers; see Fig.~\ref{fg:spheres}. (ii) In addition, the finite particle size reduces the available volume that particles can occupy, which increases the probability of a collision. Enskog accounted for this by multiplying Eq.~(\ref{molecular_chaos_2}) by a factor $\chi({\bf r}-\frac{1}{2}\sigma{\bf k})$ evaluated at the point of contact, i.e.
\be
\chi({\bf r}-\frac{1}{2}\sigma{\bf k})f_1({\bf r},{\bf v}_1,t) f_1({\bf r}-\sigma{\bf k},{\bf v}_2,t) .\label{eq:mc2}
\ee

Enskog derived his kinetic equation from similar heuristic arguments as used by Boltzmann, but replacing the molecular chaos approximation by Eq.~(\ref{eq:mc2}) and considering exclusively hard spheres. The result was $df/dt = C_\textrm{E} (f, f)$ with the collision operator
\begin{align}
\lefteqn{C_\textrm{E}(f,f) = \!\int\! d^3v^\prime d^2k \sigma^2 \vc{u} \cdot \vc{k} }&&\label{eq:ensk}\\
&&\biggl[ \chi (\vc{r} \!+\!\frac{1}{2} \sigma \vc{k}) \hat{f}(\vc{r}) \hat{f} (\vc{r} \!+\! \sigma \vc{k})- \chi (\vc{r}\! -\! \frac{1}{2} \sigma \vc{k}) f (\vc{r}) f(\vc{r} - \sigma \vc{k}) \biggr]  . \nonumber
\end{align}
Here, $\vc{u} \equiv \vc{v}_1 - \vc{v}_2$, and $\hat{f}$ denotes distribution functions evaluated at post-collision velocities $\hat{\vc{v}} = \vc{v} + \Delta \vc{v}$, whereas $f$ denotes distribution functions evaluated at pre-collision velocities $\vc{v}$. 
In order for the resulting transport equations to be consistent with the equation of state, the factor $\chi$ is taken equal to the equilibrium pair distribution function $\chi = g(\sigma/2)$ at the point of contact $\frac{1}{2}({\bf r}_1+{\bf r}_2)=\frac{\sigma}{2}{\bf k}$ between the two colliding particles.

Equation~(\ref{eq:ensk}) differs from the Boltzmann collision operator in the following ways: (a) The factor $\chi$ is absent in the Boltzmann collision operator. (b) The distribution functions are evaluated at non-local spatial locations in Eq.~(\ref{eq:ensk}), whereas all are evaluated at the local position $\vc{r}$ in the Boltzmann equation. (c) The kernel associated with the scattering probability in Eq.~(\ref{eq:ensk}), $d^2k \sigma^2 \vc{u} \cdot \vc{k}$ is particular to hard spheres, whereas in the Boltzmann equation this is replaced by $d\Omega \sigma^\prime u$ where $\sigma^\prime$ is a differential scattering cross section associated with an unspecified potential $v(r)$. 

Enskog then derived fluid transport equations by applying the Chapman-Enskog expansion to Eq.~(\ref{eq:ensk}). In comparison to the transport coefficients derived from the Boltzmann equation, aspect (a) leads to an increased collision frequency in the Enskog theory. This is manifest in the self-diffusion coefficient 
\begin{equation}
D = D_o/ \chi \label{eq:difensk} 
\end{equation}
where $D_o$ is the dilute gas self-diffusion coefficient obtained from the Boltzmann equation. For hard spheres, the lowest order expression is $D_{o,\textrm{hs}} = 3 (\pi mk_BT)^{1/2}/(8\pi mn \sigma^2)$~\cite{ferz:72}.

In addition to the increased collision frequency, the non-local aspect (b) gives rise to a non-ideal equation of state, and contribution from particle interactions in the viscosity and thermal conductivity transport coefficients. Here, we will be interested in the viscosity coefficient 
\begin{equation}
\eta = \frac{\eta_o}{\chi} [1 + 0.8 b\rho \chi + 0.7614 (b\rho \chi)^2]  \label{eq:viscensk}
\end{equation}
where $\eta_o$ is the dilute gas shear viscosity. For hard spheres, the lowest order term is $\eta_{o,\textrm{hs}} = 5(\pi mk_BT)^{1/2}/(16\pi\sigma^2)$. Here,
\begin{equation}
b\rho = \frac{2}{3} \pi n \sigma^3 
\end{equation}
is the co-volume of the molecules, which can also be expressed in terms of the packing fraction $\bar{\eta} = b \rho /4 $. The first term on the right of Eq.~(\ref{eq:viscensk}) is the kinetic term, which is the dilute gas result modified only by the increased collision probability factor $\chi$. The second two terms on the right arise due to the non-local aspects of the collision operator. These potential contributions are absent in the dilute gas theory. 

Evaluation of these coefficients requires some external determination of $\chi$. This is usually obtained by equating the equation of state implied by Eq.~(\ref{eq:ensk}), $p/(nk_BT) = 1 + b\rho \chi$, with that obtained by another means. One option is the Carnahan-Starling approximation~\cite{carn:69}, which leads to $\chi_{\textrm{CS}} = (1-\bar{\eta}/2)/(1-\bar{\eta})^3$. Similarly, the Percus-Yevick approximation~\cite{perc:58} gives $\chi_{\textrm{PY}} = (1+\bar{\eta}/2)/(1-\bar{\eta}^2)$. Another approach is to equate Enskog's equation of state with the virial coefficients of the thermodynamic equation of state for hard spheres $p/(nk_BT) =  1 + b \rho + 0.6250 (b \rho)^2 + 0.2869 (b\rho)^3 + 0.115 (b\rho)^4 + \ldots$. This implies~\cite{ferz:72}
\begin{equation}
\chi = 1 + 0.6250 b\rho + 0.2869 (b\rho)^2 + 0.115 (b\rho)^3 + \ldots  \label{eq:chi}
\end{equation}
The accuracy of these various approximations have been compared in detail (e.g., see \cite{ferz:72,hans:06}). 

It is not obvious how to extend Enskog's ansatz, Eq.~(\ref{eq:mc2}), to continuous interparticle potentials.
For instance, how does one define an effective particle size? 
The size of a hard sphere is an intrinsic property, but for continuous potentials the exclusion volume is a statistical concept and the distance of closest approach between particles in a binary collision is momentum dependent. 
Nevertheless, it is reasonable to expect that the correlation hole will affect transport properties in a similar fashion as the finite particle size of hard-spheres.
In fact, in his 1922 paper~\cite{ensk:22}, Enskog himself gave indications as to how his theory might be applied to real systems.
In the next section, we discuss challenges that arise when attempting a systematic derivation of a generalized Enskog equation for soft potentials, then we outline a phenomenological approach that enables progress on approximating fluid transport coefficients. 

\section{Modified Enskog Theory\label{sec:mensk}}

\subsection{Difficulties of a Systematic Theory}

To our knowledge, no one has yet succeeded in deriving an Enskog-like kinetic equation for continuous potentials either phenomenologically or from first-principles.
In fact, it is only in 1961 that Sengers and Cohen~\cite{seng:61} could justify the Enskog kinetic equation from the BBGKY hierarchy. They did so using two independent methods: the coarse-graining method of Bogolyubov and the time-smoothing approach of Kirkwood.
This section provides some identification of the difficulties that arise when attempting to extend such methods to treat continuous potentials.
To this end, we highlight important steps of Sengers and Cohen's derivation, giving a simplified version to make our point more transparent and accessible.
We highlight when and why extending the derivation to continuous potentials faces serious difficulties.

\begin{figure}
\centering
\includegraphics[width=8cm]{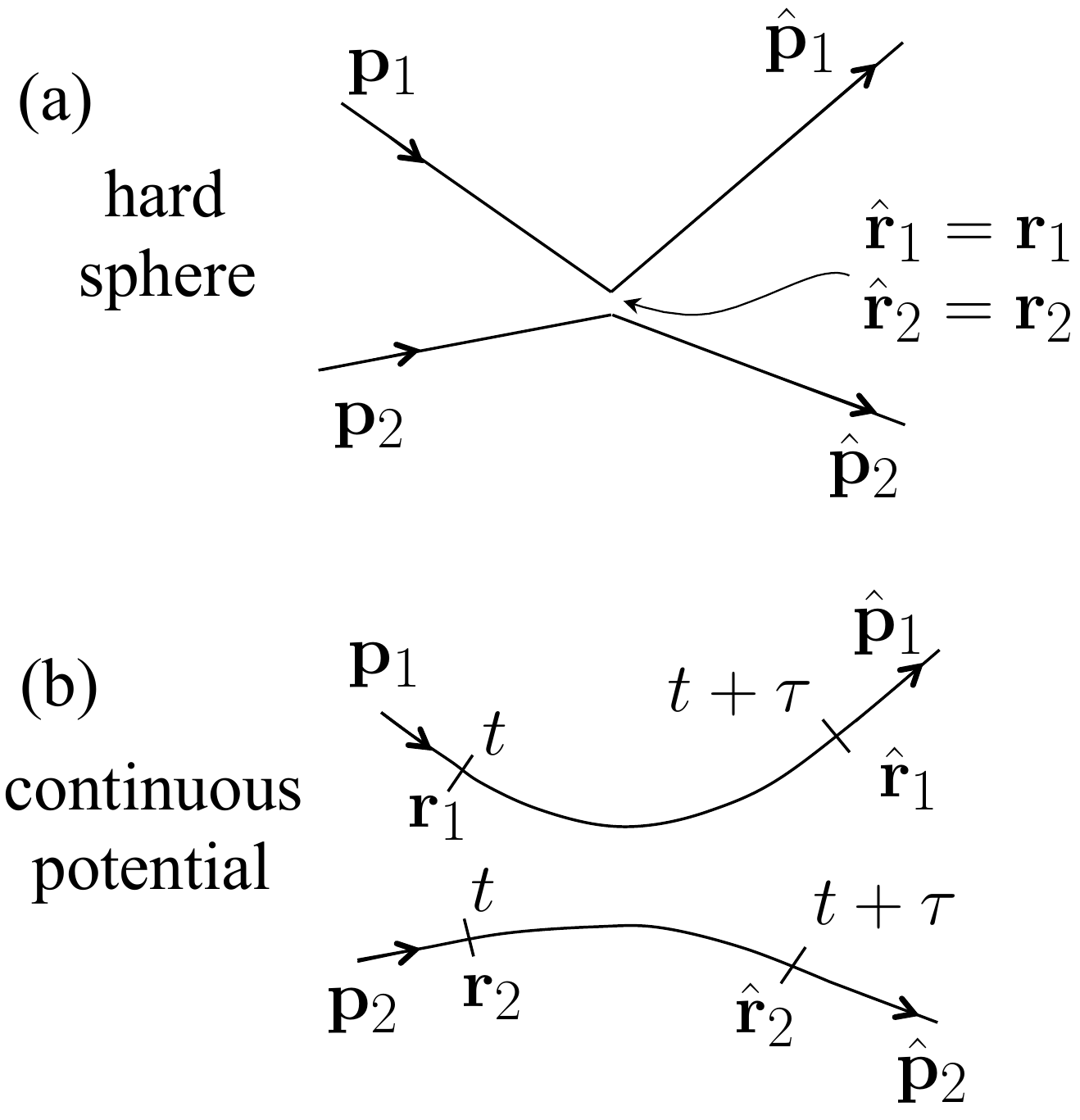}
\caption{\label{figure_collision} (a) Depiction of momentum vectors in a binary collision of hard spheres, and (b) in a binary collision of particles interacting through a continuous potential. 
}
\end{figure}
Following the coarse-graining method that Kirkwood used to derive the Boltzmann equation in his 1946 paper~\cite{kirk:46}, the distribution functions of the BBGKY hierarchy are time-smoothed over an interval from 0 to $\tau$, where the time $\tau$ is long compared to the time of a collision and short compared to the time between collisions.
For instance, for the single-particle distribution function, 
\be
\bar{f}_1(x,t)=\frac{1}{\tau}\int_0^\tau{f_1(x,t+s)ds}
\ee
where $x=({\bf r},{\bf p})$ is a point in phase-space.
The latter satisfies the first BBGKY equation,
\be
\frac{\partial \bar{f}_1}{\partial t}+{\bf v}\cdot\frac{\partial \bar{f}_1}{\partial{\bf r}}=\frac{1}{\tau}\int_0^\tau{ds\int{dx_2 L_{12} f_2(x_1,x_2;t+s)}}\equiv{\cal{C}}\nonumber\\ \label{equation_f1}
\ee
where $L_{12}=\frac{\partial \phi(r_{12})}{\partial {\bf r}_1}\cdot\left(\frac{\partial}{\partial{\bf p}_1}-\frac{\partial}{\partial{\bf p}_2}\right)$ is the interaction vertex and $f_2$ is the two-particle distribution function.
The latter satisfies the second BBGKY equation.
Thus, within the binary collision approximation underlying the effective potential theory, we can write
\be
f_2(x_1,x_2;t+s)&=&e^{-s{\cal{H}}_{12}}f_2(x_1,x_2;t) \label{equation_f2}
\ee
where ${\cal{H}}_{12}.=\{\frac{{\bf p}_1^2}{2m}+\frac{{\bf p}_2^2}{2m}+\phi(r_{12}), .\}$ is the Liouville operator for a two-particle system.
In Eq.~(\ref{equation_f2}), $e^{-s{\cal{H}}_{12}}$ propagates in time the dynamics of two particles over a time interval $\tau$.
Within this binary collision approximation, the right-hand side of Eq.~(\ref{equation_f1}) becomes
\be
{\cal{C}}=\frac{1}{\tau}\int_0^\tau{ds\int{dx_2 L_{12} e^{-s{\cal{H}}_{12}}f_2(x_1,x_2;t)}} .
\ee

For hard-spheres, all binary collisions happen instantaneously and one can take the time average over an infinitesimally small time interval $\tau\to 0^+$ during which only one binary collision can occur, and one has
\be
{\cal{C}}=\lim_{\tau\to 0+}{\frac{1}{\tau}\int_0^\tau{ds\int{dx_2 L_{12} e^{-s{\cal{H}}_{12}}f_2(x_1,x_2;t)}}} . \label{eq:cmid}
\ee
As illustrated in Fig.~\ref{figure_collision}, the particle position does not change in an instantaneous collision, so
\begin{equation}
\lim_{\tau\to 0+}e^{-s{\cal{H}}_{12}}{\bf r}_1={\bf r}_1,\quad \lim_{\tau\to 0+}e^{-s{\cal{H}}_{12}}{\bf r}_2 = \vc{r}_2 \label{position_change}
\end{equation}
whereas the momenta change discontinuously 
\begin{equation}
\lim_{\tau\to 0+}e^{-s{\cal{H}}_{12}}{\bf p}_1=\hat{{\bf p}}_1, \ \ \ \lim_{\tau\to 0+}e^{-s{\cal{H}}_{12}}{\bf p}_2=\hat{{\bf p}}_2 . \label{momentum_change}
\end{equation}
Noting that ${\bf r}_2={\bf r}_1+\sigma{\bf k}$ at contact, Eq.~(\ref{eq:cmid}) reduces to
\begin{widetext}
\be
{\cal{C}}&=&\lim_{\tau\to 0+}{\frac{1}{\tau}\int_0^\tau{ds\int{dx_2 L_{12} f_2({\bf r}_1,e^{-s{\cal{H}}_{12}}{\bf p}_1;{\bf r}_2,e^{-s{\cal{H}}_{12}}{\bf p}_2;t)}}}\\
&=&\int{d{\bf p}_2\int{d{\bf k}\sigma^2|({\bf u}\cdot{\bf k})| \left[\bar{f}_2({\bf r}_1,\hat{{\bf p}}_1;{\bf r}_1+\sigma{\bf k},\hat{{\bf p}}_2;t)-\bar{f}_2({\bf r}_1,{\bf p}_1;{\bf r}_1+\sigma{\bf k},{\bf p}_2;t)\right]}}
\ee
\end{widetext}
where the last equation is derived in~\cite{seng:61}.
Now, if in addition one assumes that $\bar{f}_2$ can be approximated by $\bar{f}_2(x_1,x_2;t)\approx g({\bf r}_1-{\bf r}_2)\bar{f}_1(x_1,t)f_1(x_2,t)$ (as in the equilibrium relation), the latter equation reduces to the Enskog kinetic equation.

The extension of the previous steps to continuous potentials leads to serious difficulties. As illustrated in Fig.~\ref{figure_collision}: (i) Collisions are not instantaneous, therefore the smoothing time scale $s$ is finite and one needs to perform the time averaging integral in ${\cal{C}}$. (ii) The duration of a collision depends on the collision parameters, therefore the time $s$ depends on collision parameters. (iii) Both positions and momenta change during a binary collision, which must be incorporated in Eqs.~(\ref{position_change})--(\ref{momentum_change}). Overcoming these difficulties is a substantial, and longstanding, challenge in kinetic theory. However, practical progress can be made through analogies between the hard-sphere system and real systems at the level of macroscopic transport equations. The remainder of this paper explores this track.

\subsection{Modified Enskog Theory For Plasmas}

Already in his seminal paper of 1922, Enskog suggested how to adjust his theory to real neutral gases in an ad-hoc manner~\cite{ensk:22}; this approach is usually called the modified Enskog theory.
The dilute gas coefficients are computed from the usual Chapman-Enskog solution of the Boltzmann equation using the interaction potential associated with the realistic system. 
In addition, he suggested that the value of $b\rho\chi$ be calculated to reproduce the variation of pressure with temperature (known as the thermal pressure) instead of simply the pressure as in the original Enskog theory.
This was found to be accurate especially for simple gases at packing fractions corresponding to $b\rho \lesssim 0.6$ (for example, see~\cite{hanl:72,parsafar:2007})

Our goal is analogous to that of the modified Enskog theories in that we seek appropriate analogies to relate key features of the realistic system to the hard sphere model.
However, a couple of difficulties are immediately apparent when the realistic system is a plasma rather than a neutral fluid. 
One is that the density is no longer the appropriate expansion parameter to link the weakly coupled to the strongly coupled regimes, and it should be replaced by the Coulomb coupling parameter.
Moreover, traditional plasma theory based on a dilute gas-like Boltzmann equation diverges in the strongly coupled regime. 
We propose that the appropriate dilute gas theory is provided by the effective potential theory from \cite{baal:13,baal:14}. Like the neutral fluid dilute gas theory, this is also based on the Boltzmann kinetic equation. The difference is that the interaction potential is the potential of mean force, rather than the bare particle potential. Making this association, the corresponding self-diffusion coefficient is
\begin{equation}
D^* \equiv \frac{D_1}{a k_BT} = \frac{1}{\chi} D_\EP^{*} .  \label{eq:densk}
\end{equation}
where $D_\EP^{*}$ is the diffusion coefficient from the effective potential theory. At lowest order, $D_{1,\EP}^{*} = \sqrt{\pi/3}/(\Gamma^{5/2} \Xi^{(1,1)})$. Here, $\Xi^{(1,1)}$ is a generalized Coulomb logarithm calculated from the effective potential theory~\cite{baal:13,baal:14}. Similarly, the shear viscosity coefficient is 
\begin{equation}
\eta^* \equiv \frac{\eta_1}{mna^2 \omega_p} = \frac{\eta_\EP^{*}}{\chi}  [ 1 + 0.8 b \rho \chi + 0.7614 (b\rho \chi)^2 ]  . \label{eq:etaensk}
\end{equation}
At lowest order, the shear viscosity coefficient obtained from the effective potential theory is $\eta_{1,\EP}^{*} = 5 \sqrt{\pi}/(3 \sqrt{3} \Gamma^{5/2} \Xi^{(2,2)})$, where $\Xi^{(2,2)}$ is a generalized Coulomb logarithm from the effective potential theory as discussed in~\cite{dali:14}.

Another apparent difficulty is that the effective size of particles is associated with a statistical ``Coulomb hole'' generated by electrostatic repulsion, rather than a hard core. Thus, a model is required to quantify the effective diameter of particles, as well as to obtain an approximation for $\chi$. Typical relationships between model and hard sphere equations of state are not viable for plasmas. For instance, the OCP pressure becomes highly negative at strong coupling due to the presence of the homogeneous neutralizing background~\cite{hans:06}.
Since both $b\rho$ and $\chi$ are positive definite quantities, the equation of state does not provide a meaningful relationship between the two systems. 
In the next section, we first determine the parameters $b\rho$ and $\chi$ in Eqs.~(\ref{eq:densk}) and (\ref{eq:etaensk}) from MD simulations of $D^*$ and $\eta^*$. The results are then compared with attempts to approximate these parameters directly from $g(r)$.

\section{Effective Size of Coulomb Point Particles\label{sec:size}}

\subsection{Constraints From MD Simulations\label{sec:sizemd}} 

If the modified Enskog theory described by Eqs.~(\ref{eq:densk}) and (\ref{eq:etaensk}) are assumed to provide an accurate model, MD simulations can be used to determine the two independent parameters $b\rho$ and $\chi$. The enhanced collision probability factor from Eq.~(\ref{eq:densk}) is 
\begin{equation}
\chi_\MD = D_\EP^*/D_\MD^* . \label{eq:chimd}
\end{equation}
Once $\chi$ is established,  $b\rho$ can then be determined from an MD computation of shear viscosity using Eq.~(\ref{eq:etaensk}) 
\begin{equation}
(b\rho)_\MD = 0.53 \chi_\MD^{-1} \biggl[ \sqrt{1 + 4.76 \biggr( \chi_\MD \frac{\eta_\MD^*}{\eta_\EP^*} -1 \biggr)} - 1 \biggr]. \label{eq:brhomd}
\end{equation}
Here $D_\MD^*$ and $\eta_\MD^*$ are the self-diffusion and shear viscosity coefficients from MD simulations. These were computed using the code and techniques described in \cite{dali:14} and \cite{dali:12}. More information is also provided in Sec.~\ref{sec:md}. 

\begin{figure}[b]
\includegraphics[width=8.0cm]{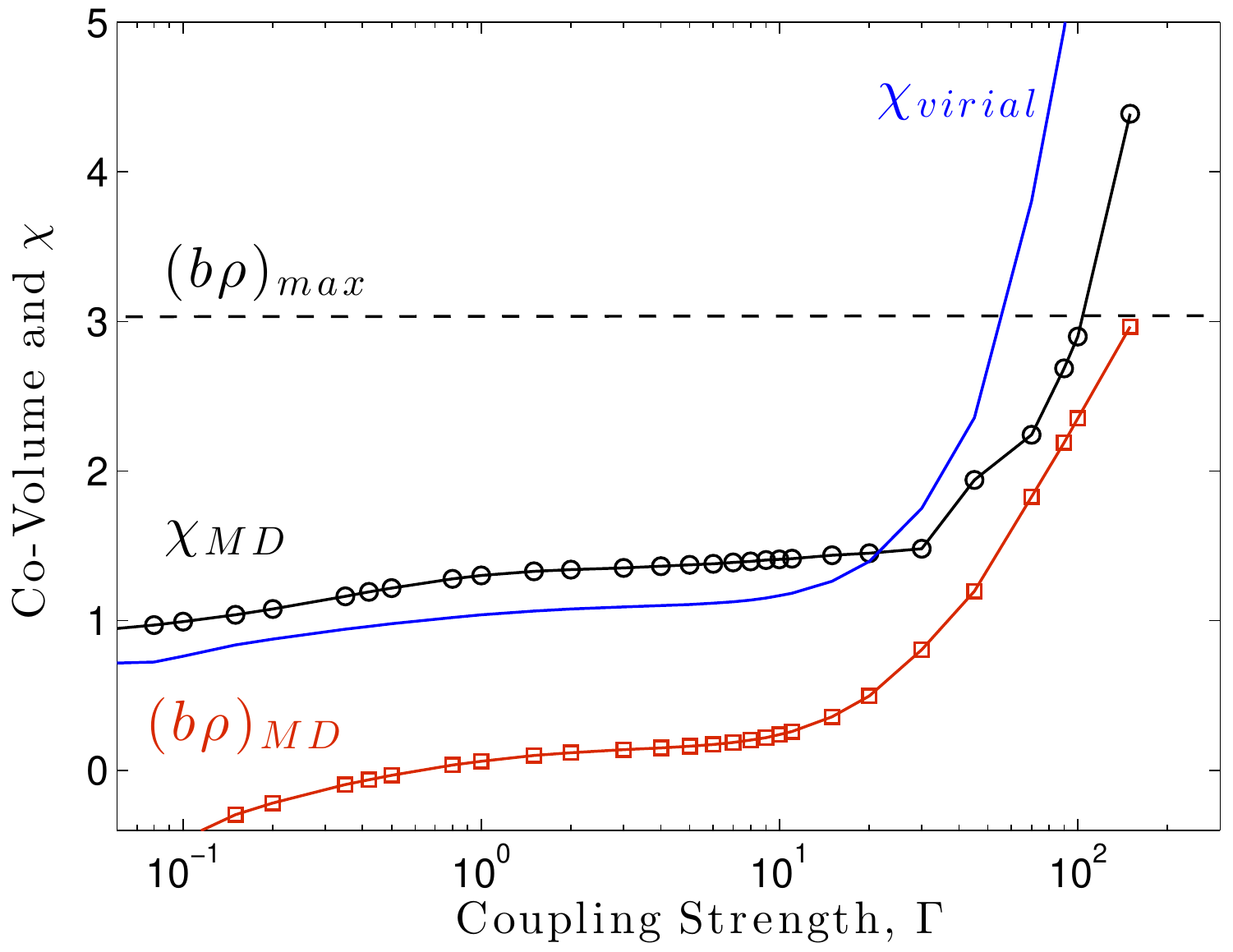}
\caption{(color online) Solutions of Eqs.~(\ref{eq:chimd}) and (\ref{eq:brhomd}) obtained using MD simulations of self-diffusivity and shear viscosity of the OCP. Also shown are the maximum physical co-volume of hard spheres (dashed line) and the virial expansion obtained using the $(b\rho)_\MD$ data in Eq.~(\ref{eq:chi}) (solid line).}
\label{fg:md_ensk}
\end{figure}

Figure~\ref{fg:md_ensk} shows the results of Eq.~(\ref{eq:chimd}) (black circles) and Eq.~(\ref{eq:brhomd}) (red squares). One apparent feature is that $\chi_\MD$ begins to differ from 1 as $\Gamma$ approaches 1. It is nearly constant over the range $1 \lesssim \Gamma \lesssim 30$, then rapidly increases for $\Gamma \gtrsim 30$. This is approximately the coupling strength at which there is a known crossover to liquid-like behavior~\cite{dali:06,dali:website}. This is discussed further in Sec.~\ref{sec:disc}. 

The MD results can be used to check that physical requirements are not violated in the modified Enskog approach. If the particle size is being modeled as an ``effective'' hard sphere, one should check that the maximum hard-sphere packing fraction is not exceeded. This is known to be $\bar{\eta}_\textrm{max} = \pi/(3\sqrt{2}) = 0.74$, which gives $(b\rho)_\textrm{max} = 4\pi/(3\sqrt{2}) = 3.0$~\cite{hans:06}. Indeed, this is approached, but not exceeded, for the range of data available in Fig.~\ref{fg:md_ensk}. It is interesting to note that the maximum physical packing fraction is being approached close to the coupling strength at which Wigner crystallization is known to occur ($\Gamma_m = 175$). The limit of the data range is set here by the maximum $\Gamma$ value at which we can obtain a numerical solution to the HNC equations. 

The MD simulations can also be used to check the virial expansion relating $b\rho$ and $\chi$. This was obtained by substituting $(b\rho)_\MD$ from Eq.~(\ref{eq:brhomd}) for the co-volume in Eq.~(\ref{eq:chi}). Figure~\ref{fg:md_ensk} shows that this is a reasonable approximation (within tens of percent accuracy) for $\Gamma \lesssim 30$, but is substantially inaccurate for larger coupling strengths. This may be expected from the fact that the virial expansion is a low-density expansion, which breaks down at sufficiently large coupling strength.

\subsection{A Heuristic Approach\label{sec:sizeh}}

As discussed before, in Enskog's theory the probability of finding a particle center within a sphere of radius $\sigma$ of another particle is zero. Here, we seek an analogy between the hard sphere system (Fig~\ref{fg:gr}a) and the Coulomb system (Fig~\ref{fg:gr}b) that provides an effective co-volume in terms of coupling strength, $b\rho(\Gamma)$, as well as the collision frequency enhancement factor $\chi(\Gamma)$.  Although the probability of finding other particles at any given separation does not completely vanish in this situation, the strong Coulomb repulsion at close distances creates a region \textit{essentially} devoid of other particles~\cite{foot1}. 

\begin{figure}
\includegraphics[width=8.0cm]{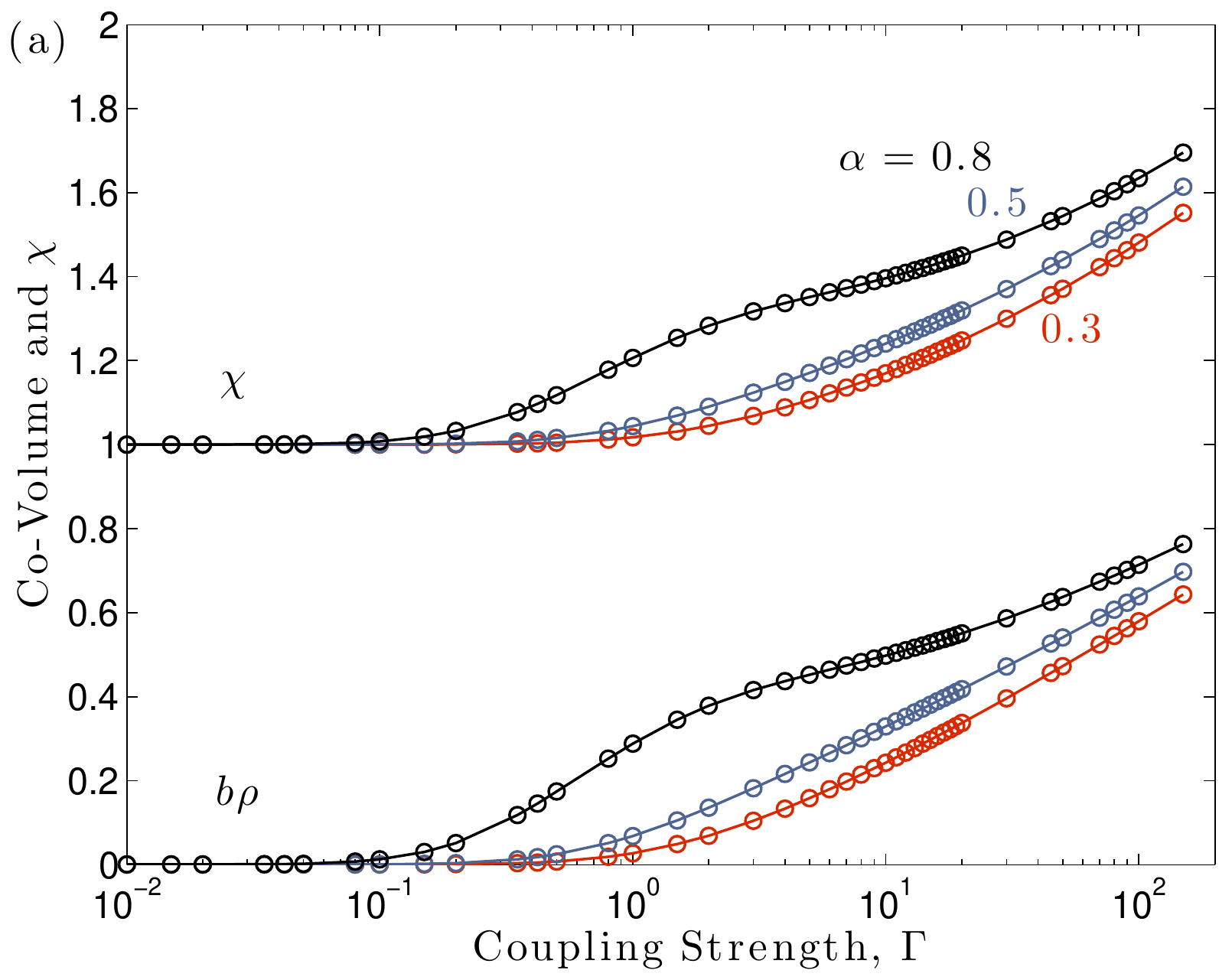}
\includegraphics[width=8.0cm]{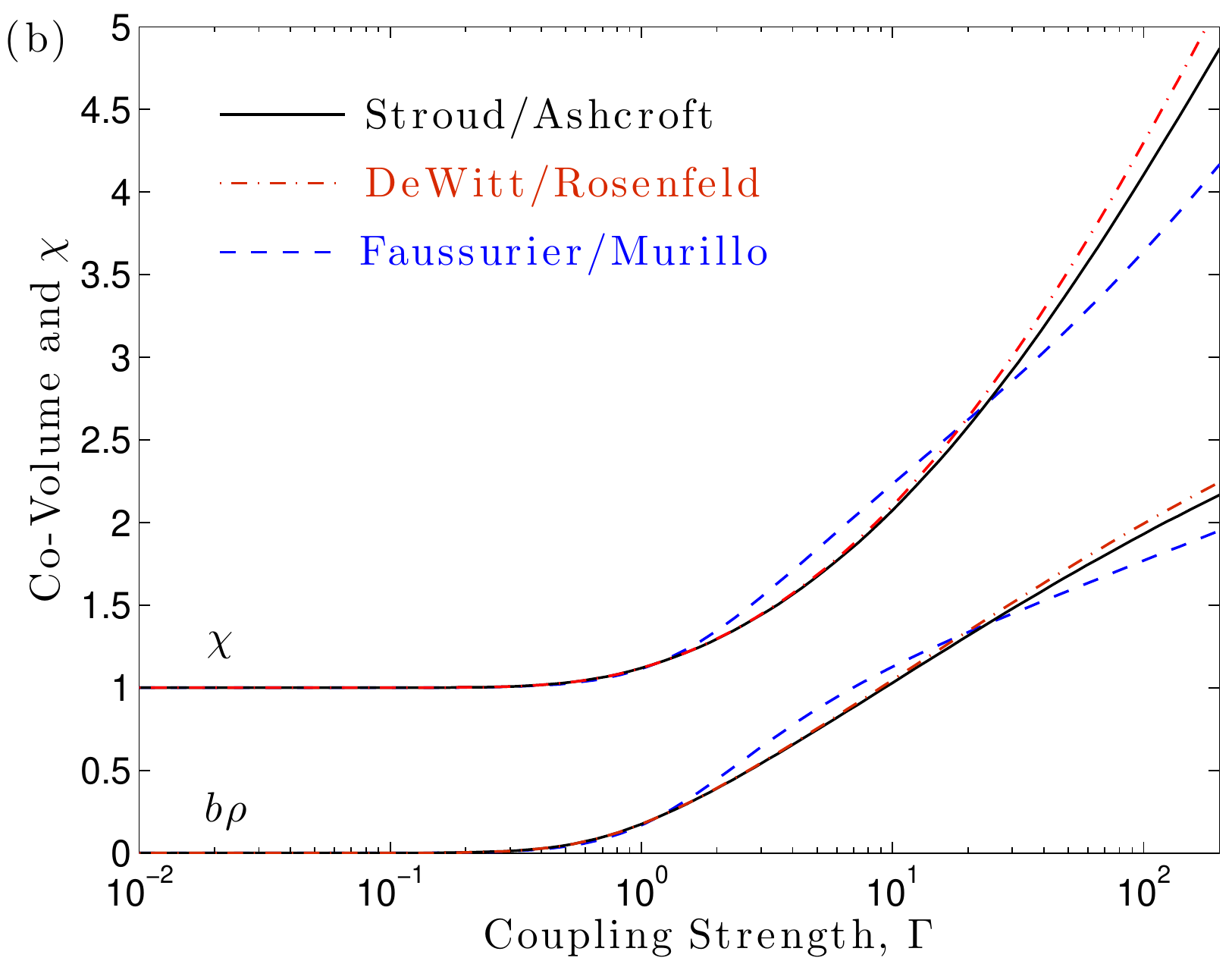}
\caption{(color online) Co-Volume ($b \rho$) and corresponding collision probably enhancement factor ($\chi$) computed, (a) using the method of Sec.~\ref{sec:sizeh} for three values of $\alpha$ and (b) using the methods of Sec.~\ref{sec:sizev}}
\label{fg:chi}
\end{figure}

To quantify this analogy, consider the pair distribution function $g(r)$. Physically, $n g(r)$ represents the radial density profile around an individual particle. Figure~\ref{fg:gr}a shows example profiles for hard spheres for three values of the packing fraction computed using the Percus-Yevick approximation~\cite{perc:58}. This is a common approximation known to accurately represent the pair distribution for the hard sphere potential. The figure shows that the density is zero within a radial distance of $\sigma$ from the particle center. The pair distribution function contains the information that particle centers must be at least a distance $\sigma$ apart because hard spheres cannot overlap. 

Figure \ref{fg:gr}b shows example profiles of the OCP pair distribution function computed from the hypernetted chain (HNC) approximation, as well as MD simulations. Like the Percus-Yevick approximation for hard spheres, HNC is a well established approximation known to accurately represent $g(r)$ for Coulomb and screened Coulomb systems~\cite{hans:06}. At weak coupling the density profile $ng(r)$ increases over a broad distance, indicating that a definite exclusion zone is difficult to identify. However, as the coupling strength increases into the strongly coupled regime, the density profile steepens substantially. Here, the exclusion radius can be identified as the radial location where $g(r)$ steps from a small value to a value of unity order.

\begin{figure}
\includegraphics[width=8.0cm]{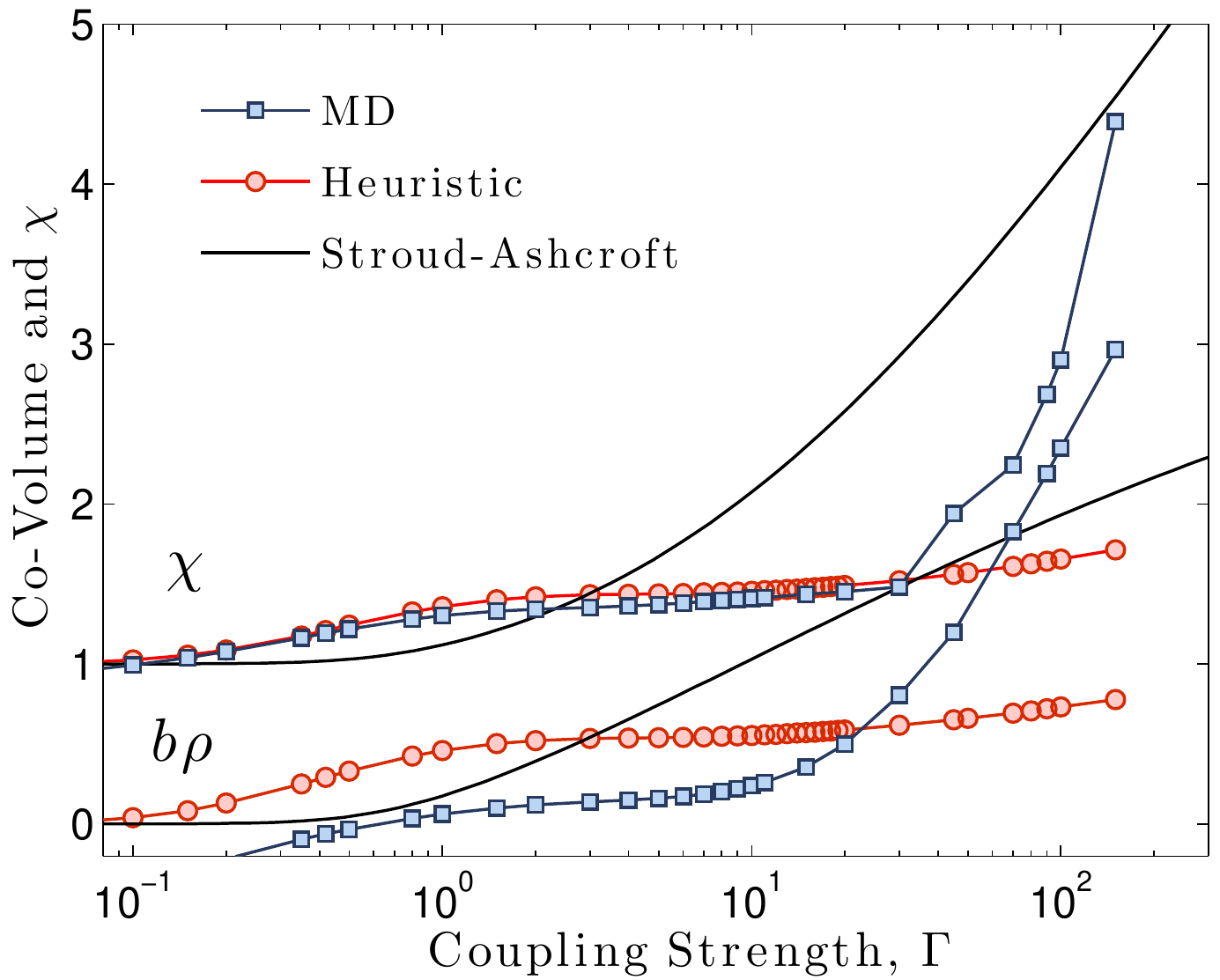}
\caption{(color online) Co-volume and collision enhancement factors computed from the MD method from Sec.~\ref{sec:sizemd} (squares), the heuristic method from Sec.~\ref{sec:sizeh} using $\alpha=0.87$ (circles) and from the Stroud-Ashcroft method from Sec.~\ref{sec:sizev} (lines).}
\label{fg:comp}
\end{figure}

The heuristic approach that we propose is to identify the location where $g(r)$ reaches some critical value ($\alpha$) as the effective particle diameter. The co-volume can then be computed from this. However, the OCP model has a uniform neutralizing background, so the co-volume associated with this diameter actually contains two particles. Thus, the individual particle co-volume is approximated as 
\begin{equation}
b\rho \simeq \frac{1}{3} \pi n \bar{\sigma}^3 = \frac{1}{4} (\bar{\sigma}/a)^3
\end{equation}
where $\bar{\sigma}$ is defined from $g(r=\bar{\sigma})=\alpha$ for some chosen value of $\alpha$. Here, we assume that $\alpha$ is constant (independent of $\Gamma$). 

For an actual hard sphere, any value in the range $0<\alpha<1$ will provide the same diameter ($\sigma$); see Fig.~\ref{fg:gr}a. For a plasma, different values of $\alpha$ will generally give different predictions for the diameter. Here, we determine the value of $\alpha$ as that which best represents the OCP data under the constraint that it be independent of $\Gamma$. Although this value will be chosen by comparison with MD data for the OCP, we hypothesize that it represents an intrinsic property of Coulomb holes for any given $g(r)$. Thus, the value of $\alpha$ is not a fitting parameter. Rather, it is presumed to be a universal value that applies to ion-ion interactions in any plasma. Although the $g(r)$ profiles will change in different systems, the critical density described by $\alpha$ will remain the same. This hypothesis will be corroborated by comparison with the Yukawa OCP model at different $\kappa$ values in Sec.~\ref{sec:mdsd}. Future work will test this hypothesis for more realistic systems. After $b\rho$ is obtained in this manner, the virial expansion from Eq.~(\ref{eq:chi}) is used to estimate $\chi$. This last step is not well justified theoretically since the expansion is for the hard sphere system, but it is this step that enables a ``mapping'' between the hard sphere model and a plasma.

Figure~\ref{fg:chi} shows the co-volume ($b\rho$) and collision probability enhancement factor ($\chi$) that result from applying this method using the HNC approximation for $g(r)$ and three different values for $\alpha$. Figure~\ref{fg:gr} shows that at weak coupling the radial density distribution gradually increases from 0 to 1, so an effective exclusion radius is not a well-defined concept in this regime. However, Fig.~\ref{fg:chi} shows that at weak coupling the co-volume is very small anyway, so although the expected error in exclusion radius might be large it has a negligible affect on the transport quantities in this regime. As the coupling strength increases into the moderately coupled regime, $\Gamma \gtrsim 0.1$, the radial density step becomes steeper and the co-volume becomes finite. This leads to values of $\chi$ that give rise to order unity corrections to the transport coefficients. 

Figure~\ref{fg:comp} shows that the heuristic method agrees very well with the MD evaluation of $\chi$ for all $\Gamma \lesssim 30$ when the value $\alpha=0.87$ is chosen. The agreement for co-volume is less accurate, but is of the order expected from the virial expansion (see Fig.~\ref{fg:md_ensk}). The heuristic method is entirely unable to capture the regime $\Gamma \gtrsim 30$. This may demonstrate the limitations of a dense gas picture, since it is known that liquid-like behaviors such as caging set in this regime~\cite{dali:06}. This is discussed further in Sec.~\ref{sec:disc}. 

\subsection{Relation to Other Works\label{sec:sizev}}


Previous work has also considered a correspondence with hard sphere systems to gain insight into plasma behavior. Ross and Seale~\cite{ross:74} sought a mapping between the OCP Coulomb coupling parameter ($\Gamma$) and the hard sphere packing fraction ($\bar{\eta}$) using a variational method. The basis of this approach is the Gibbs-Bogolyubov inequality, which relates the Helmholtz free energy of a given system to that of a reference system, which in this case is the hard sphere gas. This was used to approximate thermodynamics quantities with some success, especially at large coupling strengths ($\Gamma \gtrsim 100$) for the OCP. A few different results based on this method have been published. Stroud and Ashcroft~\cite{stro:76} obtained the relation
\begin{equation}
\Gamma = 2 \bar{\eta}^{1/3} \frac{ 2 - \bar{\eta}}{2 + \bar{\eta}} \frac{(1 + 2 \bar{\eta})^2}{(1 - \bar{\eta})^5}  . \label{eq:sa}
\end{equation}
This analysis used the Percus-Yevick $g(r)$ to obtain the internal energy, and the Carnahan-Starling expression~\cite{carn:69} to obtain the hard sphere entropy. DeWitt and Rosenfeld~\cite{dewi:79} provided an alternative expression
\begin{equation}
\Gamma = 2 \bar{\eta}^{1/3} \frac{(1 + 2 \bar{\eta})^2}{(1- \bar{\eta})^4} \label{eq:dr}
\end{equation}
which uses the Percus-Yevick $g(r)$ to obtain both the internal energy and hard sphere entropy. More recent results by Faussurier and Murillo~\cite{faus:03} provide the expression
\begin{align}
& \ln (\bar{\eta}) = 3 \ln \Gamma - \ln (8) + \label{eq:fm} \\ \nonumber
& [1.845 + 0.006 \ln \Gamma] \ln \biggl\lbrace \frac{\exp(-1.503 \ln \Gamma + 0.22)}{1 + \exp (-1.503 \ln \Gamma + 0.22)} \biggr\rbrace 
\end{align}
which uses an improved Percus-Yevick approximation for the internal energy along with the Carnahan-Starling expression for the hard sphere entropy. 

\begin{figure}
\includegraphics[width=8.0cm]{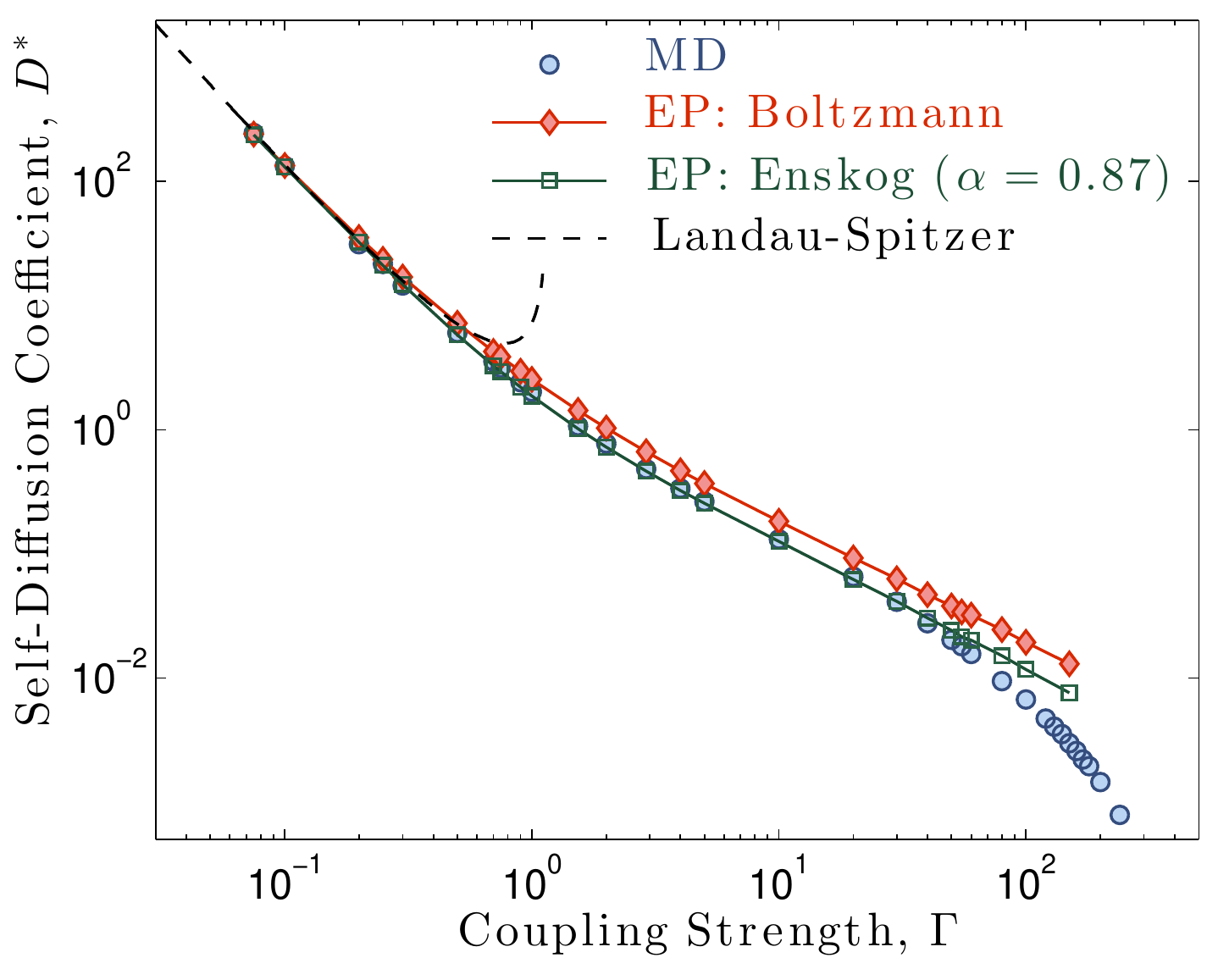}
\caption{(color online) Normalized self-diffusion coefficient of the OCP computed from MD simulations (circles), the effective potential theory based on the Boltzmann-like collision operator (triangles), and the modified Enskog method from Sec.~\ref{sec:sizeh} using $\alpha = 0.87$ (squares).}
\label{fg:selfd}
\end{figure}

The results of the co-volume obtained from Eqs.~(\ref{eq:sa})--(\ref{eq:fm}) are shown in Fig.~\ref{fg:chi}b. The $\chi$ factor obtained from applying these in Eq.~(\ref{eq:chi}) are also shown. The results of this method lead to a much larger estimated packing fraction at strong coupling than the method of Sec.~\ref{sec:sizeh}. Stroud and Ashcroft have shown that the variational upper bound of the excess free energy associated with this approach agrees well with Monte Carlo simulation results only at very strong coupling ($\Gamma \gtrsim 100$); see Fig.~2 of \cite{stro:76}. Thus, it is likely that the heuristic method underestimates the packing fraction at very strong coupling, but the variational method overestimates it for $\Gamma \lesssim 100$. Figure~\ref{fg:comp} shows that these theories do not accurately capture the MD evaluation of the Enskog theory parameters over the range of coupling strengths were the effective potential approach applies.

\section{Transport Coefficients\label{sec:md}}

\subsection{Self-Diffusion\label{sec:mdsd}} 

Figure~\ref{fg:selfd} shows a comparison between MD simulations and the effective potential theory for the OCP self-diffusion coefficient. The MD simulation data was computed from the Green-Kubo relation as explained in~\cite{dali:12}. The theory was computed two ways. The first method evaluated the usual effective potential theory as described in \cite{baal:13}. This applies the potential of mean force computed from the HNC approximation to a Boltzmann kinetic equation and the associated diffusion coefficient from the Chapman-Enskog solution. This corresponds to $D_\EP^{*}$ in Eq.~(\ref{eq:densk}), and includes up to the second order in the Chapman-Enskog expansion; see Eq.~(14) of \cite{baal:13} for the formula. The second method included the Enskog correction, $D_\EP^{*}/\chi$ [see Eq.~(\ref{eq:densk})]. Here, the factor $\chi$ was computed using the method of Sec.~\ref{sec:sizeh} with the value $\alpha = 0.87$. 

\begin{figure}
\includegraphics[width=8.0cm]{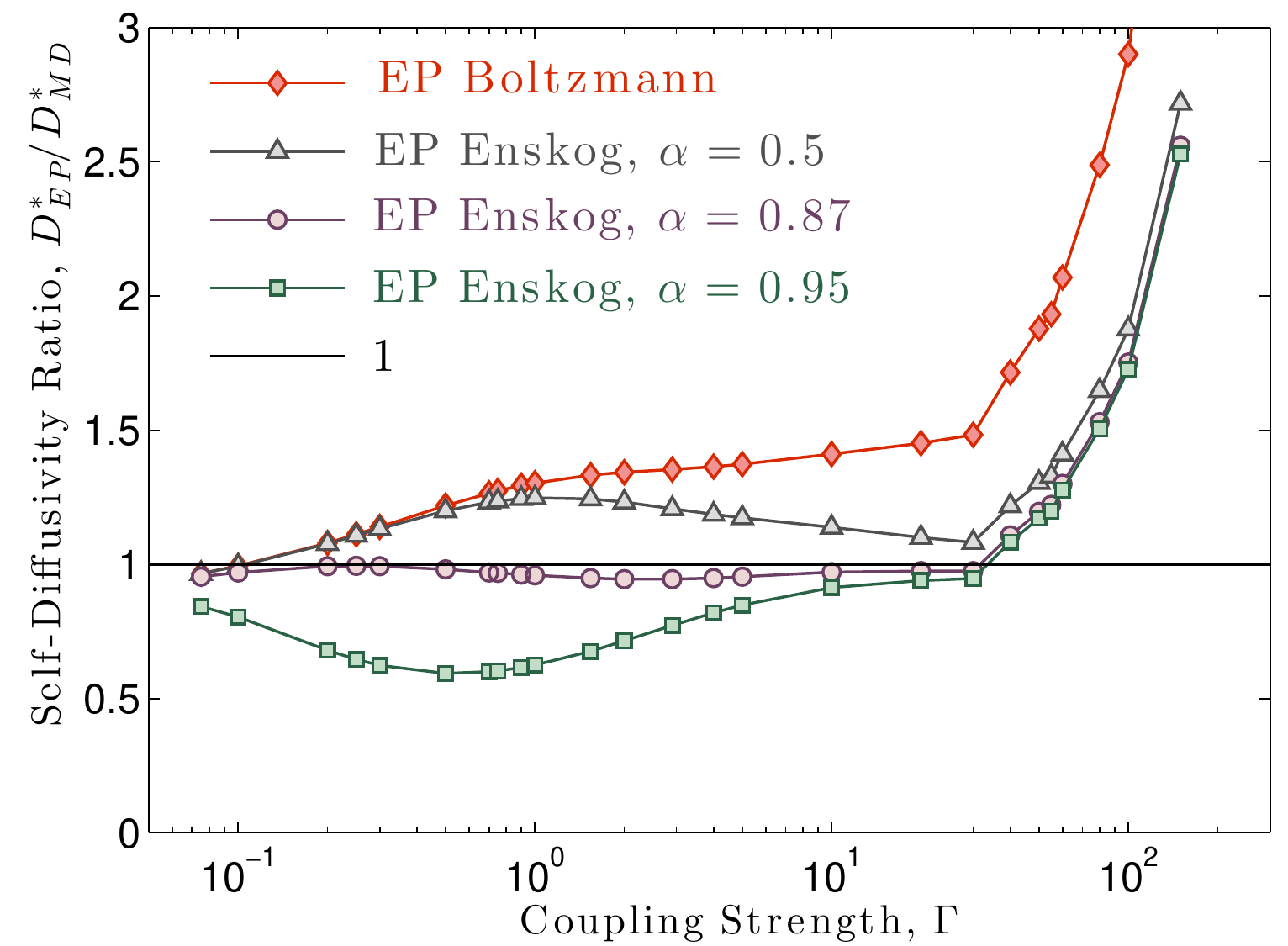}
\caption{(color online) Ratio of the OCP self diffusion coefficients calculated using MD simulations and the effective potential theory based on the Boltzmann-like collision operator (diamonds) and including the Enskog correction based on the method of Sec.~\ref{sec:sizeh} for $\alpha = 0.5, 0.87$ and 0.95. }
\label{fg:selfd_r}
\end{figure}

\begin{figure}
\includegraphics[width=8.0cm]{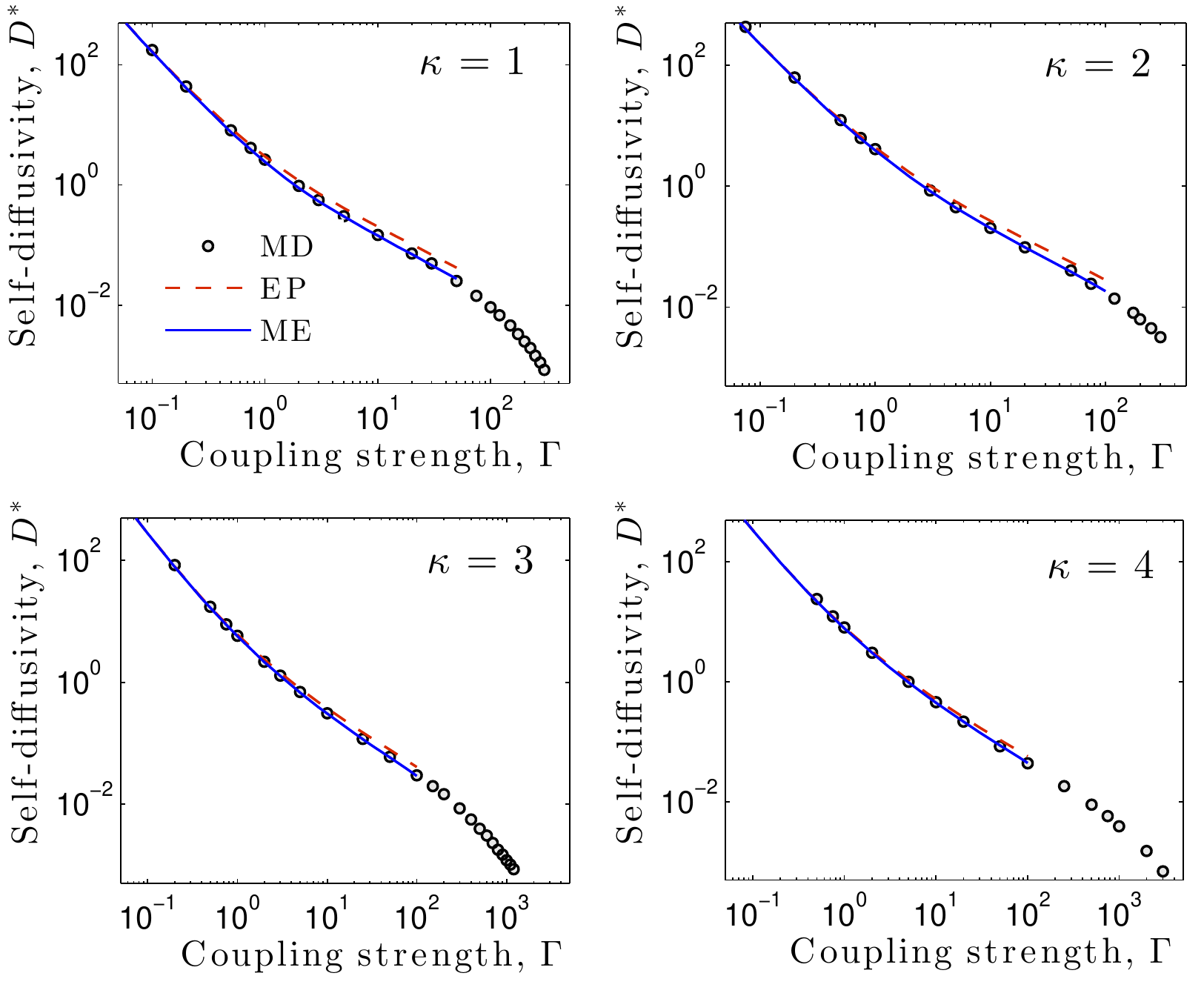}
\caption{(color online) Self-diffusion coefficient for the Yukawa OCP at four different screening parameters calculated using MD (circles), the effecitive potential theory (dashed red lines), and the modified Enskog method from Sec.~\ref{sec:sizeh} using $\alpha = 0.87$ (solid blue lines). }
\label{fg:ensk_d}
\end{figure}

\begin{table}
\caption{Values of the self-diffusion coefficient computed from MD simulations and Eq.~(\ref{eq:densk}) where the second order Chapman Enskog expansion is used to calculated $D_{\EP}^*$.}
\begin{center}
\begin{tabular}{l  c  c  c | c  c c c }
\hline \hline
$\Gamma$   &  $D_{\textrm{MD}}^*$  & $D_2^*$   &  \% diff \ \  & \ \ $\Gamma$   &  $D_{\textrm{MD}}^*$  & $D_2^*$   &  \% diff     \\ \hline
0.075     & 246     & 237     &  3.4   & 2.9  & 0.482 & 0.464 & 3.8    \\ 
0.1         & 133       & 131     &  1.5   & 4.0  & 0.337 & 0.324 & 3.9    \\ 
0.2         & 31.3    & 32.5     &  3.7   & 5.0  & 0.264 & 0.255 & 3.4   \\ 
0.25     & 21.6     & 21.0     &  2.6   & 10  & 0.131 & 0.125 & 3.9    \\ 
0.3     & 14.5     & 14.8     &  2.1   & 20  & 0.0653 & 0.0622 & 4.8    \\ 
0.5     & 6.02     & 5.81     &  3.5   & 30  & 0.0411 & 0.0414 & 0.6   \\ 
0.7     & 3.56     & 3.28     &  7.8   & 40  & 0.0275 & 0.0303 & 10   \\ 
0.75     & 3.16     & 2.93     &  7.2   & 50  & 0.0202 & 0.0241 & 19    \\ 
0.9     & 2.41     & 2.20     &  8.7   & 55  & 0.0180 & 0.0215 & 20    \\ 
1.0     & 2.02    & 1.87    &  7.3   & 60  & 0.0156 & 0.0201 & 30   \\ 
1.54     & 1.07     & 1.02     &  5.1   & 80  & 0.00944 & 0.0151 & 60    \\ 
2.0     & 0.770     & 0.72     &  6.1   & 100  & 0.00672 & 0.0117 & 74    \\ 
\hline \hline
\end{tabular}
\end{center}
\label{tb:diff}
\end{table}

Including Enskog's collision probability enhancement factor ($\chi$) provides a substantial improvement to the effective potential theory. A subset of the data points shown in the figure are given in table~\ref{tb:diff} along with the \% difference calculated as $|D_{\textrm{MD}}^* - D_\EP^*/\chi|/D_\textrm{MD}^*$.  Figure~\ref{fg:selfd_r} shows the ratio of the theoretical and MD results. Without the Enskog correction, the effective potential theory overestimates the self-diffusion coefficient by approximately 30-40\% in the range $1 \lesssim \Gamma \lesssim 30$. Enskog's $\chi$ factor essentially corrects this error if the effective exclusion radius is chosen from an appropriate value of $\alpha$. Here $\alpha = 0.87$ provides good agreement over the coupling parameter range of interest. Other values of $\alpha$ also lead to an improved theory, but a value near 0.87 provides the most accurate fit. 

The $\alpha$ factor is obtained directly from $g(r)$ based on the concept that there is a critical density that defines the particle radius. Thus, the numerical value is expected to be system-independent. This notion is supported by Fig.~\ref{fg:ensk_d}, which shows the self-diffusion coefficient for the Yukawa OCP model at four different screening parameters $\kappa = 1, 2, 3$ and 4. Here, the modified Enskog approach from Sec.~\ref{sec:sizeh} is shown to provide a similar accuracy for the Yukawa OCP as the OCP when the same value $\alpha = 0.87$ is chosen. 

\subsection{Shear Viscosity} 

\begin{figure}
\includegraphics[width=8.0cm]{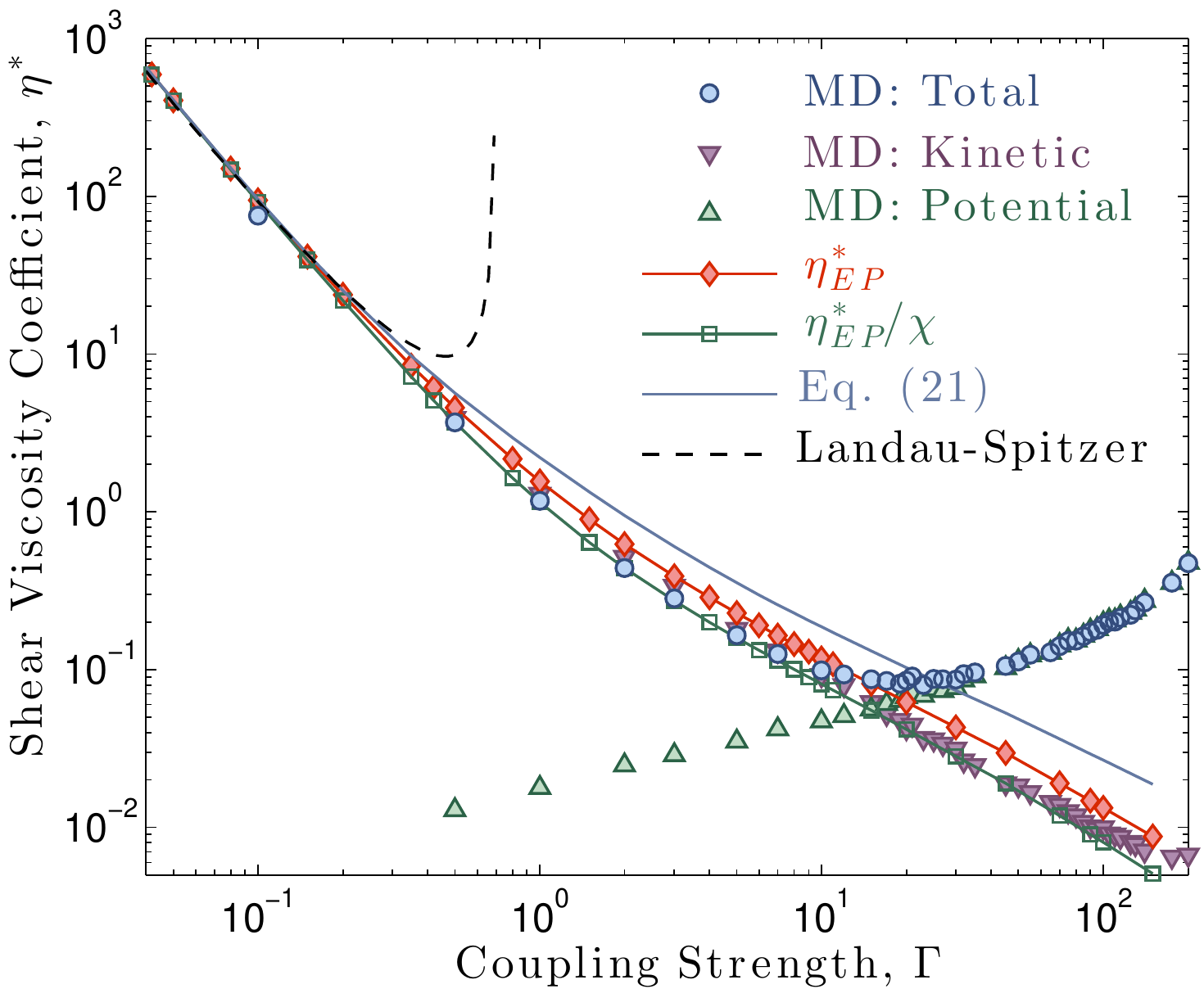}
\caption{(color online) Normalized shear viscosity of the OCP computed from MD: total (circles), kinetic contribution (downward triangles) and potential contribution (upward triangles). Three theoretical evaluations are shown: $\eta_\EP^*$, $\eta_\EP^*/\chi$ and Eq.~(\ref{eq:etaensk}), for which the heuristic method with $\alpha = 0.87$ was used to determine $\chi$ and $b\rho$.  }
\label{fg:visc}
\end{figure}

Figure~\ref{fg:visc} shows a comparison between MD simulations and the EP theory for shear viscosity of the OCP. Here, three different methods for evaluating the EP theory are shown. One is the previous evaluation from \cite{dali:14} based on the Boltzmann collision operator ($\eta_\EP^{*}$). Here the coefficient associated with the first order Chapman-Enskog expansion is shown (the formula is given after Eq.~(\ref{eq:etaensk}) above). The second is a simple modification that treats the lowest-order increased scattering probability factor ($\eta_\EP^{*}/\chi)$, but not the potential terms in Enskog's theory that arise from the non-local feature of the collision operator. The third evaluation includes the complete Enskog expression from Eq.~(\ref{eq:etaensk}). In addition, the partial contributions from kinetic and potential terms of viscosity are shown individually from the MD data; details are provided in \cite{dali:14}. 

The figure shows that, once again, application of Enskog's increased scattering probability factor ($\chi$) improves the effective potential theory over the range $1 \lesssim \Gamma \lesssim 30$. Applying this correction leads to excellent agreement between the theory and kinetic component of the MD simulations over the entire range of coupling strengths shown. However, the full modified Enskog solution, including the nonlocal effects arising from the $b\rho$ terms in Eq.~(\ref{eq:etaensk}), significantly degrades the accuracy of the theory. Furthermore, the method appears to be incapable of capturing the minimum in viscosity. The minimum occurs where the potential components exceed the kinetic components. These effects are meant to be modeled by the terms in square brackets in Eq.~(\ref{eq:etaensk}), however the large $\Gamma$ at which this transition occurs coincides with where the heuristic approach for obtaining the particle size breaks down.

\section{Discussion of results\label{sec:disc}} 

The results of sections \ref{sec:size} and \ref{sec:md} substantiate the supposition that a modified Enskog theory can be used to improve the effective potential transport theory, but there are limitations. There is particularly strong evidence that the increased collision frequency ($\chi$) associated with the reduced volume that particles of a finite size can occupy leads to a significant improvement in the range $1\lesssim \Gamma \lesssim 30$.  Although there is some arbitrariness in how the effective particle size is defined (the $\alpha$ factor) even a simple heuristic approach essentially removed any discrepancy between the theoretical and MD calculations in this regime. It is also supportive that any physically reasonable value of $\alpha$ led to an improvement of the theory, although some values were found to lead to better results than others.
In a future publication, we will test the validity of the present theory against molecular dynamics simulations of realistic plasmas using the techniques recently presented in \cite{starrett:2015}

Although the modified Enskog theory led to a substantial improvement in the range $1 \lesssim \Gamma \lesssim 30$, it essentially failed at modeling the potential contributions to viscosity that arise for $\Gamma \gtrsim 30$. One might suggest that this is due to an inaccurate model for the effective particle size in this regime. However, if this is the explanation, Fig.~\ref{fg:md_ensk} shows that the effective particle size must make an abrupt adjustment to a much steeper scaling with $\Gamma$ to extend Enskog's arguments to this regime. It is hard to imagine on a physical basis why the effective particle diameter should increase so abruptly at $\Gamma \simeq 30$ if the medium is to remain dense gas-like. Instead, this may be indicative of the breakdown of the dense gas-like picture and the onset of a liquid-like regime. 

The data presented here gives further evidence of the following analogy between transport regimes of the OCP and neutral fluids: (i) Dilute gas like regime ($\Gamma \lesssim 1$). This is supported by the finding that the dilute-gas Boltzmann equation, using an effective interaction potential, accurately predicts transport coefficients in this regime. (ii) Dense gas like regime ($1 \lesssim \Gamma \lesssim 30$). Here a Boltzmann equation with effective interaction potential gives reasonable results, but these are improved by also including dense gas effects. In particular, the finite effective size of particles gives rise to a nonnegligible increase in the collision frequency. (iii) Liquid like regime ($30 \lesssim \Gamma \lesssim 175$). Although there is no abrupt gas-liquid phase transition for the OCP, as there is in neutral fluids, there is substantial evidence for liquid-like behavior in this regime. Previous work has shown that particle caging dominates transport, and that liquid-state scaling relationships, such as Stokes-Einstein, the Arrhenius law of viscosity, and other excess-entropy scaling relationships hold in this regime~\cite{donk:02,dali:06}. Additionally, the present work has shown that an Enskog-type dense gas theory abruptly runs into difficulties as this regime is approached. (iv) Crystalline regime ($\Gamma \gtrsim 175$). The OCP undergoes a well-known phase transition to a crystal lattice near $\Gamma = 175$. 


\begin{acknowledgments}

The authors thank Drs. Saumon and Starrett for helpful conversations that motivated this work, and Prof.~E.G.D. Cohen for helpful conversations on Enskog kinetic theory. This research was supported under the auspices of the National Nuclear Security Administration of the U.S. Department of Energy at Los Alamos National Laboratory under Contract No. DE-AC52-06NA25396.


\end{acknowledgments}


\begin{thebibliography}{99}


\bibitem{baal:13} S.~D.~Baalrud and J.~Daligault, Phys.~Rev.~Lett.~{\bf 110}, 235001 (2013).

\bibitem{baal:14} S.~D.~Baalrud and J.~Daligault, Phys.~Plasmas {\bf 21}, 055707 (2014).

\bibitem{bezn:14} M.\ V.\ Beznogov, and D.\ G.\ Yakovlev, Phys.\ Rev.\ E {\bf 90}, 033102 (2014).

\bibitem{hanl:72} H.~J.~M.~Hanley, R.~D.~McCarty and E.~G.~D.~Cohen, Physica {\bf 60}, 322 (1972).

\bibitem{ensk:22} D.~Enskog, Kungl.~Svenska Vet.-Ak.~Handl.~{\bf 63} (1922).

\bibitem{ferz:72} J.\ H.\ Ferziger and H.\ G.\ Kaper, \emph{Mathematical theory of transport processes in gases}, (Elsevier, New York, 1972).

\bibitem{hill:60} T.\ L.\ Hill, \emph{An Introduction to Statistical Thermodynamics} (Addison-Wesley, Reading, 1960) p.\ 313.

\bibitem{seng:61} J.~V.~Sengers and E.~G.~D.~Cohen, Physica {\bf 27}, 230 (1961).

\bibitem{carn:69} N.~F.~Carnahan and K.~E.~Starling, J.~Chem.~Phys.~{\bf 51}, 635 (1969).

\bibitem{perc:58} J.~K.~Percus and G.~J.~Yevick, Phys.~Rev.~{\bf 110}, 1 (1958). 

\bibitem{hans:06} J.-P.\ Hansen and I.\ R.\ McDonald, \emph{Theory of Simple Liquids, 3rd Edition} (Academic Press, Oxford, 2006).

\bibitem{kirk:46} J.~G.~Kirkwood, J.~Chem.~Phys.~{\bf 14}, 180 (1946). 

\bibitem{parsafar:2007} G.A. Parsafar and Z. Kalantar, Fluid Phase Equilibrium {\bf 253}, 108 (2007).

\bibitem{dali:14} J.~Daligault, K.~\O.~Rasmussen and S.~D.~Baalrud, Phys.~Rev.~E {\bf 90}, 033105 (2014).

\bibitem{dali:12} J.~Daligault, Phys.~Rev.~Lett.~{\bf 108}, 225004 (2012).

\bibitem{dali:06} J.~Daligault, Phys.~Rev.~Lett.~{\bf 96}, 065003 (2006). 

\bibitem{dali:website} Illustrative movies showing the gas-like to liquid-like crossover can be viewed at this webpage:  http://www.lanl.gov/projects/dense-plasma-theory/research/one-component-plasma.php

\bibitem{foot1} In this work we consider only the OCP. Treatment of attractive collisions is not immediately apparent within this picture. 

\bibitem{ross:74} M.~Ross and D.~Seale, Phys.~Rev.~A {\bf 9}, 396 (1974).

\bibitem{stro:76} D.~Stroud and N.~W.~Ashcroft, Phys.~Rev.~A {\bf 13}, 1660 (1976). 

\bibitem{dewi:79} H.~E.~DeWitt and Y.~Rosenfeld, Phys.~Lett.~{\bf 75A}, 79 (1979).

\bibitem{faus:03} G.~Faussurier and M.~S.~Murillo, Phys.~Rev.~E {\bf 67}, 046404 (2003).

\bibitem{donk:02} Z.\ Donk\'{o}, G.\ J.\ Kalman and K.\ I.\ Golden, Phys.\ Rev.\ Lett.\ {\bf 88}, 225001 (2002). 

\bibitem{starrett:2015} C. E. Starrett, J. Daligault, and D. Saumon, Phys. Rev. E {\bf 91}, 013104 (2015).



\end{thebibliography}

\end{document}